\documentclass[12pt,preprint]{aastex}
\begin{document}
\title{The Anisotropy in the Galaxy Velocity Field originated from the 
Gravitational Pancaking Effect}
\author{Yookyung Noh and Jounghun Lee}
\affil{Department of Physics and Astronomy, FPRD, Seoul National
University, Seoul 151-747, Korea} 
\email{ykyung@astro.snu.ac.kr,jounghun@astro.snu.ac.kr}

\begin{abstract}
We analyze the Millennium run semi-analytic galaxy catalog to explore  
quantitatively the gravitational pancaking effect on the orientation 
of galaxy velocity field.  We first calculate the probability density 
distribution of the cosine of the angle between the velocity of a field 
galaxy and the direction normal to a local pancake plane which is determined 
using two nearest neighbor field galaxies. A clear signal of alignment is 
detected for the case that the pancake scale is in the range of $5-8h^{-1}$ Mpc.  
The tendency of the velocity-pancake alignment is found to still exist when 
the pancakes are determined using three neighbor galaxies, indicating 
that it has a spatial coherence. The degree of the velocity-pancake alignment 
is shown to increase with the velocity magnitude and the local density, while it 
decreases with the separation distance from  the galaxy to the pancake and 
disappears when the pancake has a filamentary shape. A final conclusion is that 
our work may provide another clue to understanding the large-scale structure 
in the universe. 
\end{abstract}

It is an undeniable fact that there are large-scale anisotropies in the universe. 
In the context of the cold dark matter (CDM) paradigm, the large-scale anisotropies are 
understood as a nonlinear manifestation of the primordial tidal field \citep{bon-etal96}.

The large-scale anisotropy induced in the distribution 
and the orientation of galaxies has been extensively studied both observationally and 
theoretically \citep{fli-god86,bar-efs87,gar-etal93,lee04,nav-etal04,tru-etal06}. But, 
its presence in the galaxy velocity field has not attracted enough attention so far 
due to the difficulty in measuring accurately the galaxy velocities from observations.

The standard theory of structure formation based on the CDM paradigm, however, 
makes a unique prediction for the anisotropy in the galaxy velocity field.
The large-scale coherence in the initial tidal field leads to the formation of 
two-dimensional sheet-like objects, pancakes \citep{sha-etal95,pau-mel95,bon-etal96}.
It will in turn lead the galaxy velocity field to possess planar symmetry, which is 
referred to as the {\it gravitational pancaking effect}. 

The pancakes are in fact characteristic of the Zel'dovich approximation \citep{zel70}, 
according to which the trajectory of a cosmic particle in the comoving coordinate is given as 
\begin{equation}
\label{eqn:zel} 
{\bf x}={\bf q}-D(t)\nabla\Psi({\bf q}).
\end{equation}
Here, ${\bf q}$ and ${\bf x}$ are the Lagrangian and the Eulerian coordinates, 
respectively,  $D(t)$ is the linear density growth factor, and  $\Psi({\bf q})$ 
is the linear velocity potential. This Zel'dovich approximation breaks down after 
the moment of the first caustics corresponding to the formation of pancakes. 
The limitation of the Zel'dovich approximation may be overcome by truncating the 
nonlinear scale powers,  which amounts to smoothing the velocity potential on the 
galaxy scale \citep{pau-mel95} . Equation (\ref{eqn:zel}) thereby is valid for 
describing the dynamical path of a galaxy till the moment of the pancake formation.

Taking the time derivative of equation (\ref{eqn:zel}) and Taylor-expanding 
$\Psi({\bf q})$ to first order, we have 
\begin{equation}
\label{eqn:shear}
\dot{x}_{i}\approx -\dot{D}\partial_{i}\Psi({\bf q})
\propto q_{j}T_{ij}, \qquad T_{ij}\equiv\partial_{i}\partial_{j}\Psi .
\end{equation}
where $T_{ij}$ is the gravitational tidal shear tensor. Since the gravitational collapse to form a pancake 
occurs along the major principal axis of the tidal shear tensor which is in the direction normal to the 
plane of the pancake, equation (\ref{eqn:shear}) predicts the alignment between ${\bf v}$ and the 
direction normal to the plane of the pancake.

We test this analytic prediction of the Zel'dovich approximation against N-body simulations 
by analyzing the galaxy catalog from the Millennium run simulation for the concordance 
$\Lambda$CDM cosmogony \citep{spr-etal05}. The catalog contains total $9925229$ galaxies 
in a periodic box of size $500h^{-1}$Mpc with information on various galaxy properties such as 
position, velocity, mass, and so on.  For our purpose, only field galaxies are chosen from the 
catalog, since for the cluster galaxies the following contaminations may occur in reality: 
First, for the cluster galaxies the dominant gravitational force would degrade the pancaking 
effect. Second, if cluster galaxies were contained within spherical virialized objects with 
tangential velocity dispersion larger than the radial one, the same kind of anisotropy would 
be misinterpreted as a sign of the pancaking effect. 

Total $854738$ field galaxies are identified from the catalog according to the criterion suggested 
by \citet{ela-pir97} and \citet{hoy-vog02} that a field galaxy should have less than three neighbors 
within a spherical radius of $d+3\sigma_{d}/2$ ($d$ is the mean distance to the third 
nearest neighbor and $\sigma_{d}$ is the standard deviation on $d$). 

To examine the velocity-pancake alignment, it is necessary first to identify pancakes from the catalog. 
Since there is no well-defined pancake-identification method, we propose the following 
expedient scheme: For each field galaxy, we find two nearest neighbor field galaxies and define a 
local pancake plane encompassing the field galaxy and the two neighbors.

Let ${\bf R}_{1}$ and ${\bf R}_{2}$ represent the displacement vectors to the first and the second nearest 
neighbors from the position of a given field galaxy, respectively, and let $R_{c}$ represent a distance lower-limit 
satisfying $R_{c} \le R_{1} < R_{2}$ where $R_{1} \equiv \vert {\bf R}_{1}\vert$ and $R_{2}\equiv \vert{\bf R}_{2}\vert$. 
We measure the cosine of the relative angle, $\theta$, between the galaxy velocity vector, ${\bf v}$, and the direction 
normal to the pancake plane as
\begin{equation}
\cos\theta \equiv \frac{\vert{\bf v}\cdot({\bf R}_{1}\times{\bf R}_{2})\vert}
{\vert{\bf v}\vert\vert{\bf R}_{1}\times{\bf R}_{2}\vert}, 
\end{equation}
and determine the probability distribution, $p(\cos\theta)$, statistically.  
We repeat this calculation with changing the value of $R_{c}$. 

\begin{figure} \begin{center} \epsscale{0.8} \plotone{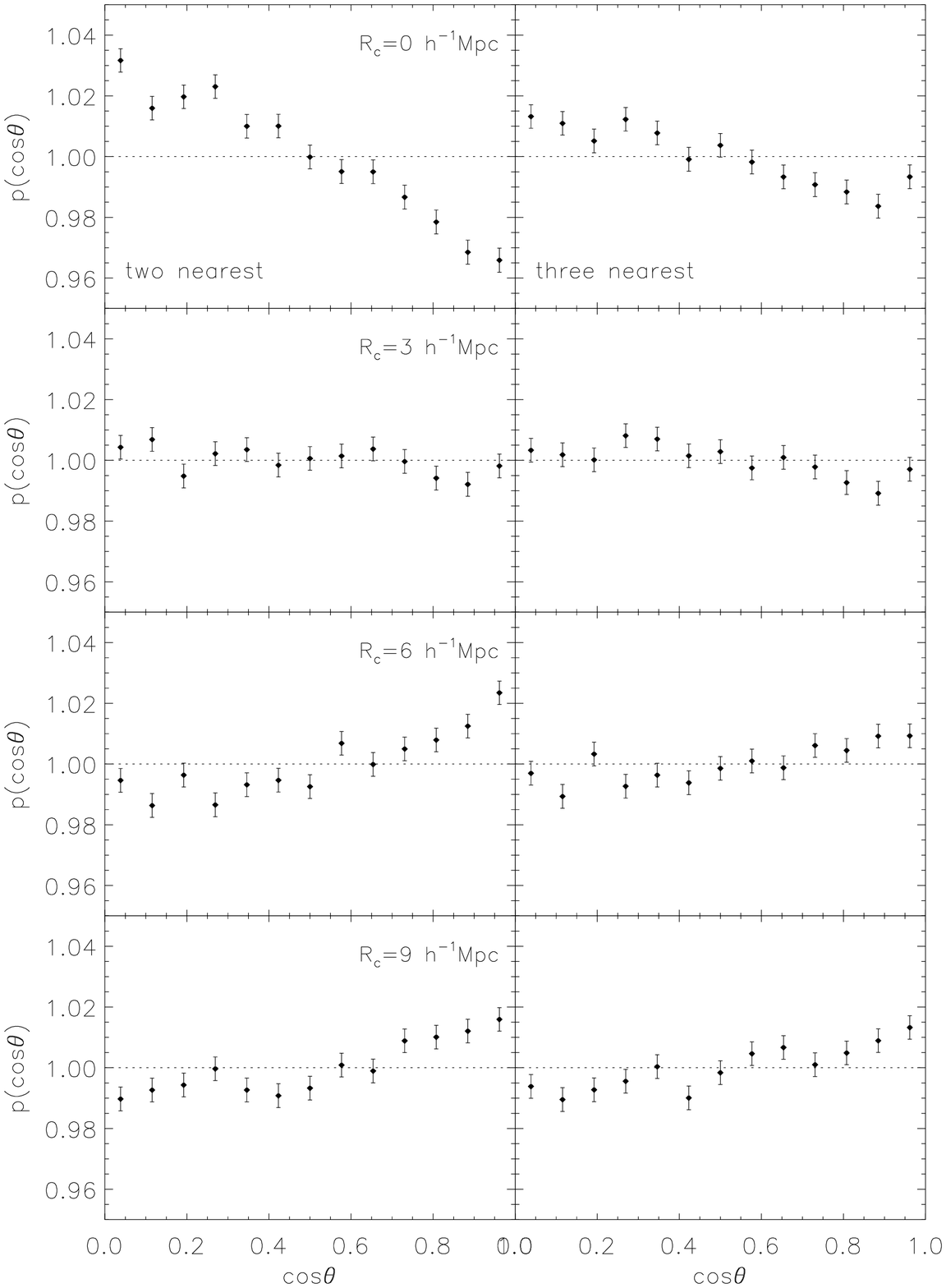}
\caption{Probability density distributions of the cosines of the relative
angles between the velocities of the field galaxies and the directions
normal to the local pancakes for the cases of the distance threshold: 
$R_{c}=0,3,6,9 h^{-1}$Mpc (from the upper to the lower panels). 
The data is taken from the Millennium run semi-analytic galaxy catalog \citep{spr-etal05}. 
The left four panels correspond to the case that a pancake is determined as a plane encompassing 
a given galaxy and its two nearest neighbors, while the right four panels correspond to the 
case that a pancake for a given galaxy determined as a plane encompassing its three nearest 
neighbors. The errors are Poissonian and the horizontal dotted line corresponds to the case 
of no alignment in each panel.}
\label{fig:pro}
\end{center}
\end{figure}

Figure \ref{fig:pro} plots $p(\cos\theta)$ as dots with the Poissonian errors for the cases 
of $R_{c}=0,3,6$ and $9$ in unit of $h^{-1}$Mpc in the four left panels, respectively.
The horizontal dotted line in each panel represents the case of no alignment. As one can see, 
for the case of $R_{c} < 3h^{-1}$Mpc,  the galaxy velocities tend to be anti-aligned 
with the directions normal to the pancakes. This tendency can be understood as due to the 
attractive forces exerted among  very close neighbors on one another. For the case of $R_{c}= 3 h^{-1}$Mpc 
the anti-alignment tendency almost disappears and $p(\cos\theta)$ looks like an uniform distribution 
having no alignment.  When $R_{c} \ge 6h^{-1}$Mpc,  obvious signals of velocity-pancake 
alignments appear, which are found to be significant at $99.9\%$ confidence level. 

\begin{figure} \begin{center} \epsscale{1.0} \plotone{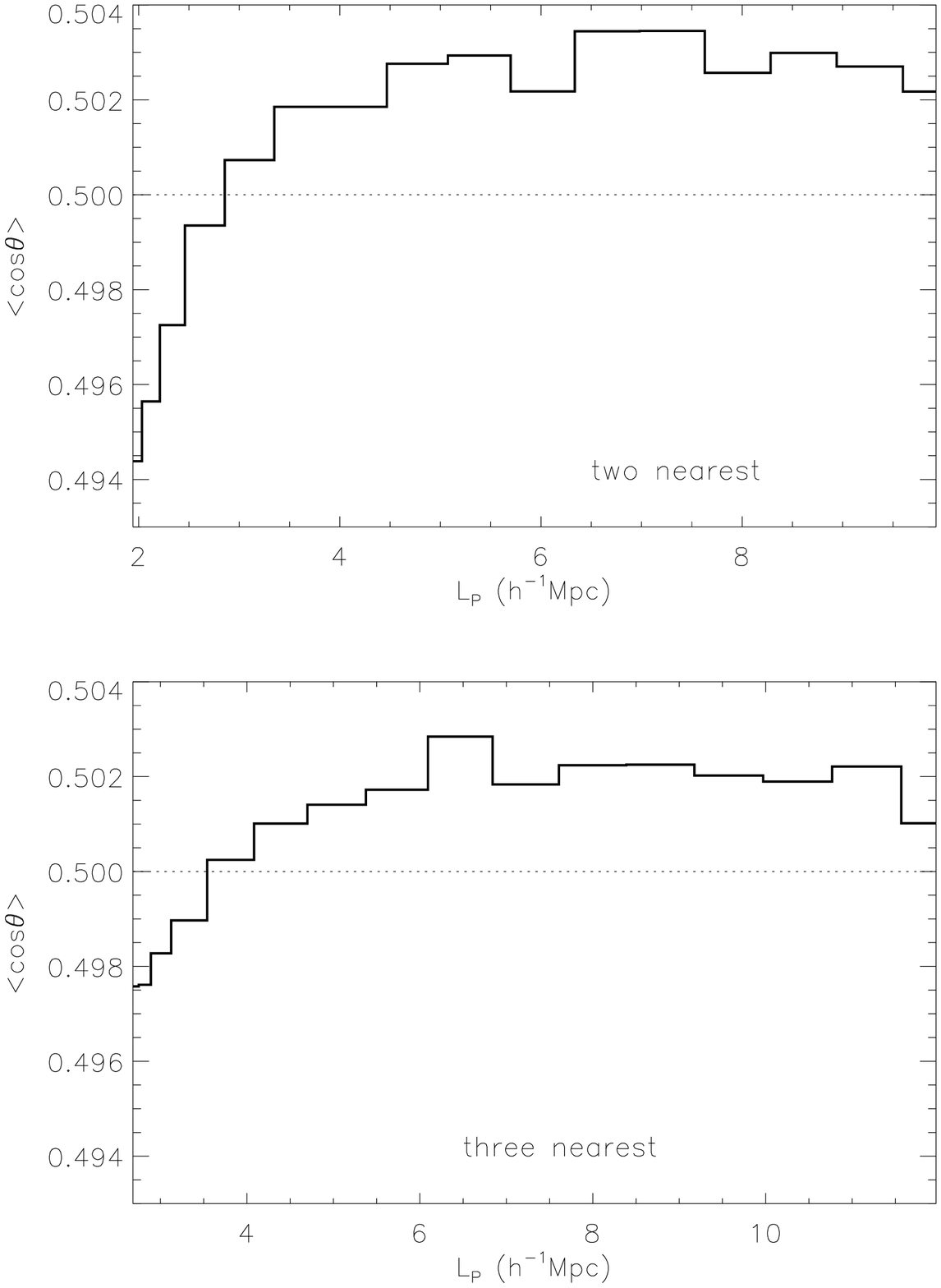}
\caption{Average of the cosines of the alignment angles as a function
of the linear size of the pancake, $L_{p}$. The upper panel corresponds to the case 
that a pancake is determined as a plane encompassing the given galaxy and the two neighbor 
galaxies, while the lower panel to the case that a pancake for a given galaxy defined as its 
three neighbor galaxies.The horizontal dotted line corresponds to the case of no alignment 
in each panel. }
\label{fig:his}
\end{center}
\end{figure}

We find the center of mass (CM) for the given and the two neighbor galaxies and compute 
the mean value of the distances from CM to the three galaxies. The pancake scale, $L_{p}$, is 
defined as this mean distance. 
We then calculate the average of the cosine of the alignment angle, $\langle\cos\theta\rangle$, 
as a function of $L_{p}$. Figure \ref{fig:his} plots the result as histogram in the upper panel. 
The horizontal dotted line represents the case of no alignment $\langle\cos\theta\rangle =0.5$. 
As can be seen, the pancaking effect on the galaxy velocity field exists when the value of $L_{p}$ lies  
in the range of $5-8h^{-1}$Mpc and gradually dies off at $L_{p} \ge 10h^{-1}$Mpc. 
We refer to this range of $L_{p}=5-8h^{-1}$Mpc where the pancaking effect is significant 
as the characteristic pancake scales .

Now that we have found a clear signal of the pancaking effect on the galaxy velocity field, 
it will be interesting to test whether the alignment tendency still exists or not when a local 
pancake is determined  using {\it three} nearest neighbor field galaxies. Note that 
if the pancake is determined in this way, then the given galaxy does not necessarily lie on the 
plane of the pancake. 

The results are shown in the four right panels of Fig. \ref{fig:pro}.
Obviously, a very similar tendency of the velocity-pancake alignment is detected for this case, too. It implies that 
that the signal of velocity-pancake alignment is robust, having a spatial coherence. 
We also recalculate $\langle\cos\theta\rangle$ as a function of $L_{p}$ for this case, the result of which is 
plotted as histogram in the lower panel of Fig.\ref{fig:his}. As can be seen, the characteristic pancake scale 
becomes slightly larger, $L_{p}=6-10h^{-1}$Mpc for this case. 

\begin{figure} \begin{center} \epsscale{0.8} \plotone{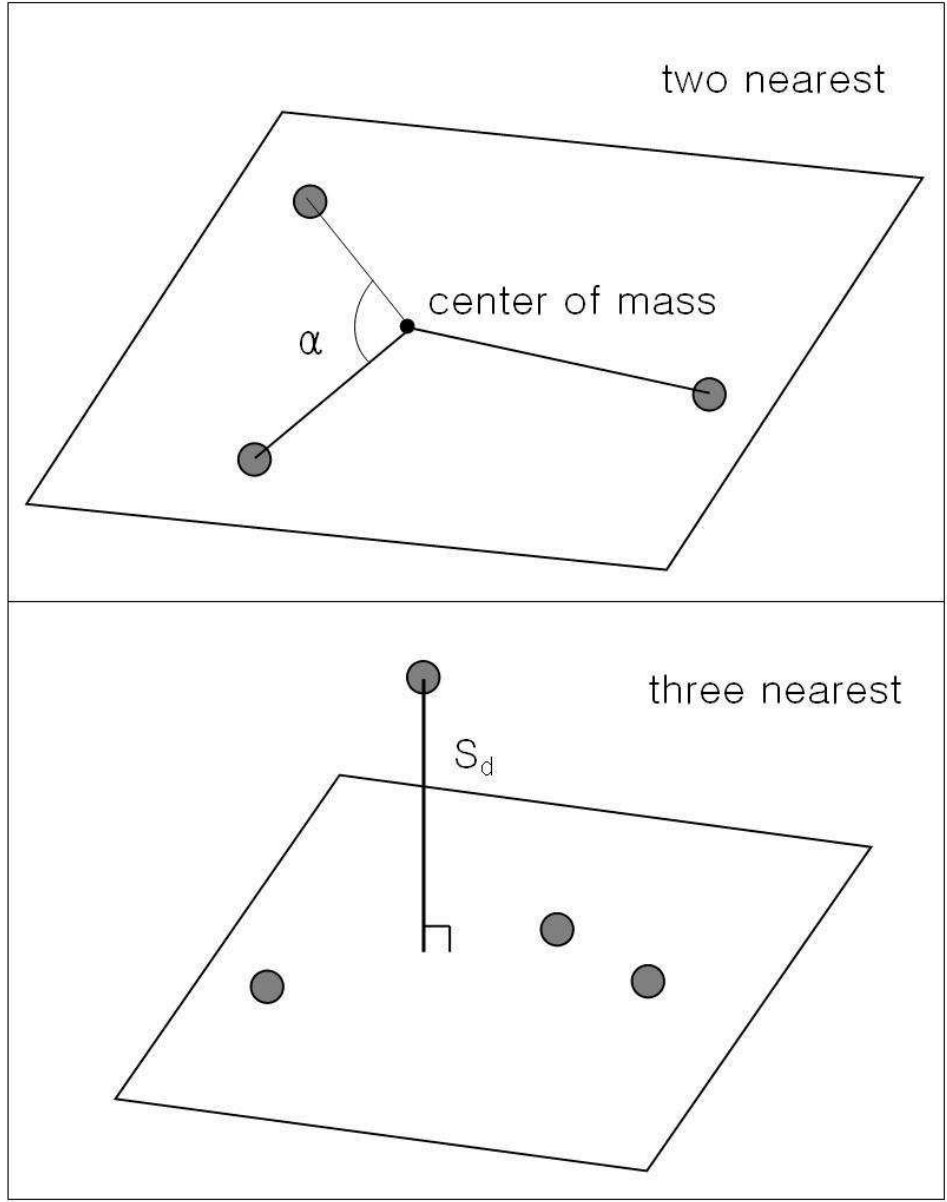}
\caption{The upper panel illustrates the configuration of the relative
angle,$\alpha$, when the local plane is defined as a plane encompassing the given 
galaxy and its two neighbor galaxies. The lower panel panel depicts the value 
of $S_{d}$which is the distance between the given galaxy and the pancake plane 
enclosing three neighbor galaxies.}
\label{fig:con}
\end{center}
\end{figure}

Using those galaxy-pancake pairs which show the alignment tendency, we investigate 
the dependence of the pancaking effect on the magnitude of galaxy's velocity ($v$), 
the local number density of galaxies ($N_{g}$), the pancake's planarity ($\cos\alpha$), and the 
galaxy-pancake distance ($S_{d}$). Here, the value of $N_{g}$ is obtained by counting the 
number of all galaxies (both field and cluster) inside a fixed radius of $6h^{-1}$Mpc. 
And, the value of $S_{d}$ represents the separation distance from the given galaxy to 
the pancake plane that is determined using the three nearest neighbors.  The lower panel of 
Fig. \ref{fig:con} illustrates how $S_{d}$ is defined. 

The upper panel of Fig. \ref{fig:con} illustrates the configuration of the three galaxies on the pancake 
plane and how the pancake planarity is defined. Among the three relative angles made by the three galaxy 
position vectors from the CM, the angle $\alpha$ is chosen as the one whose cosine has the smallest 
absolute value.  Note that the pancake planarity is lowest if $\cos\alpha \approx 1$, which in fact 
corresponds to the case of filamentary shape. 

\begin{figure} \begin{center} \epsscale{1.0} \plotone{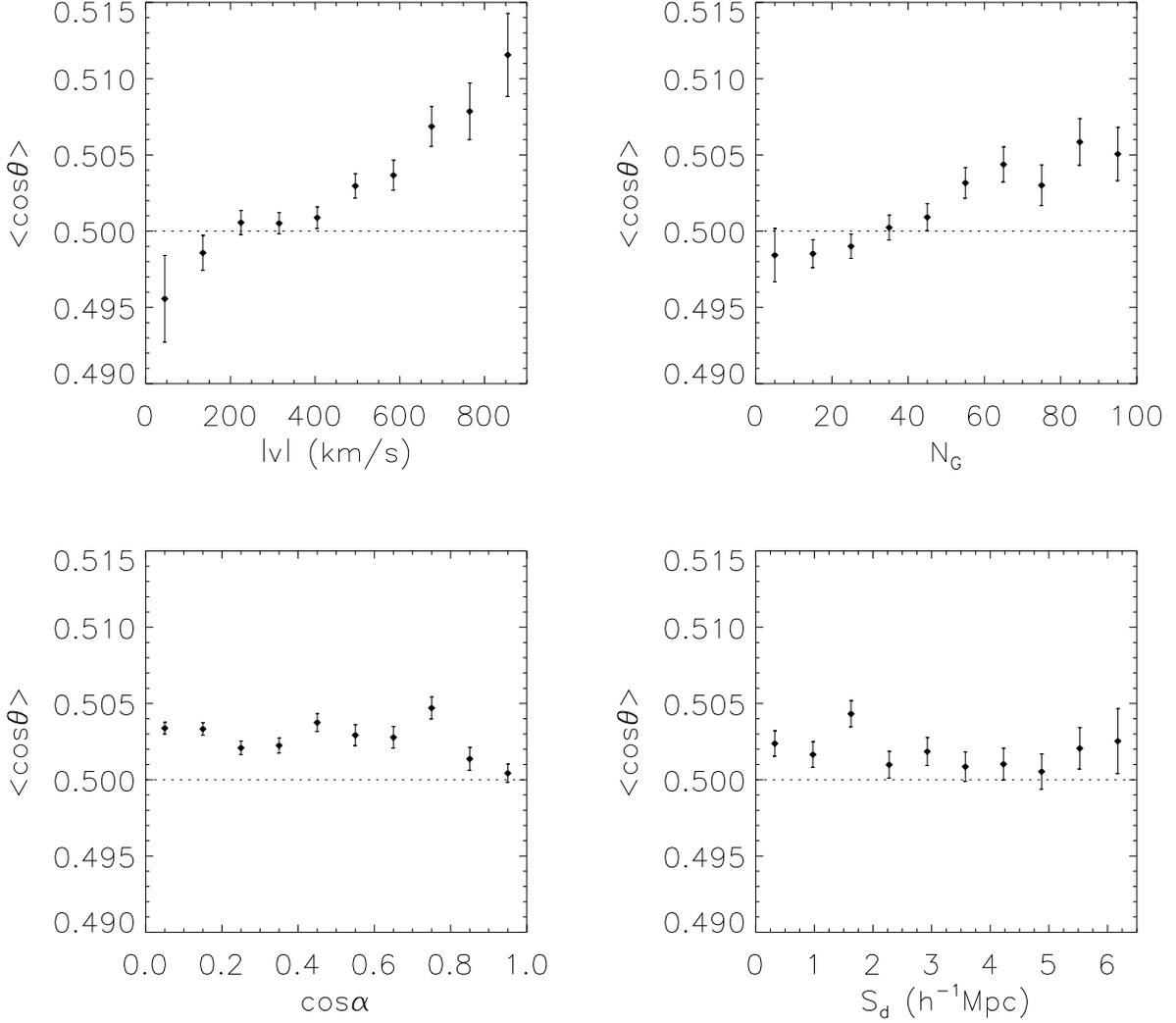}
\caption{Average of the cosines of the alignment angles as a function
of the galaxy's velocity magnitude (upper left), number density of galaxies
within the distance of $6h^{-1}$Mpc (upper right), the largest cosines of angles
between the neighbor galaxies (lower left), and the distance between
the galaxy and the pancake plane (lower right). The horizontal dotted line
in each panel corresponds to the case of no alignment.}
\label{fig:ave}
\end{center}
\end{figure}

We calculate $\langle\cos\theta\rangle$ as a function of $v$, $N_{g}$, $\cos\alpha$ and 
$S_{d}$, the results of which are plotted as dots with errors in the upper left, upper right, lower left, 
and lower right panels of Fig. \ref{fig:ave}, respectively. The errors here are computed as the standard 
deviation for the case of no correlation: 
$\sqrt{(\langle\cos^{2}\theta\rangle-\langle\cos\theta\rangle^{2})/N}$ 
where $\langle\cos^{2}\theta\rangle=1/3$, $\langle\cos\theta\rangle=1/2$, and $N$ is the number of galaxies 
in each bin. As can be seen, the degree of the velocity-pancake alignment increases sharply with $v$ and $N_{g}$, 
indicating those galaxies with high velocity located near massive pancakes experience stronger pancaking effect. 

As can be seen in the lower left panel of Fig.\ref{fig:ave}, the velocity-pancake alignment effect 
disappears for the case that the pancakes have very low planarity (i.e., $\cos\alpha \sim 1$).  
The lower right panel of Fig. \ref{fig:ave} reveals another intriguing result that the strongest 
signal of the velocity-pancake alignment effect is attained when the distance to the pancake, $S_{d}$, 
is approximately $2h^{-1}$Mpc. Note, however, that compared with the results shown in the other panels 
the actual signal that attains a maximum at $S_{d}\sim 2h^{-1}$Mpc is much smaller.
 
Now, we would like to discuss the difficulty of detecting the pancaking effect in practice. 
The signal of the velocity-pancake alignment effect is revealed to be very low. Observational detection 
of it would require very accurate measurement of the positions and the velocities of millions of galaxies 
and thus may not be achievable from the currently available data. Notwithstanding, by developing a new 
statistical quantification of the pancaking effect on the galaxy velocity field, it might be possible to 
observe the effect. Our future project is in this direction and hope to report the result somewhere else 
in the future.

As final conclusions, we have, for the first time, investigated quantitatively the anisotropy 
in the galaxy velocity field and found that the galaxy velocities have a tendency to be aligned in 
the directions normal to the local pancakes of characteristic scales of $5-8h^{-1}$Mpc. 
Our work may provide a deeper insight to the intrinsic properties of the large-scale structure 
in the universe.

\acknowledgments 
We thank the anonymous referee who helped us improve significantly the 
original manuscript. The Millennium Run simulation used in this paper was carried out 
by the Virgo Supercomputing Consortium at the Computing Centre of the Max-Planck 
Society in Garching. The semi-analytic galaxy catalogue is publicly available at 
\\ http://www.mpa-garching.mpg.de/galform/agnpaper.
J.L. acknowledges stimulating discussion with E.L. Turner.
This work is supported by the research grant No. R01-2005-000-10610-0 from 
the Basic Research Program of the Korea Science and Engineering Foundation.

\clearpage
\clearpage
\clearpage
\clearpage

\end{document}